\documentclass[twocolumn,showpacs,amsmath,amssymb,prl]{revtex4}

\usepackage{psfrag}
\usepackage{graphicx}%
\usepackage{dcolumn}%
\usepackage{bm}%
\usepackage{epsfig}
\usepackage{yfonts}
\usepackage{version}
\usepackage[utf8]{inputenc}

\usepackage{gensymb}  %

\usepackage{color} 
\usepackage{ulem}  %
\newcommand{\bra}[1]{\langle\,{#1}\, |}
\newcommand{\ket}[1]{|\,{#1}\,\rangle}

\newcommand{\Real}{\mbox{Re}}

\newcommand{\ElTransE}{ \varepsilon}

\newcommand{\LichtFr} { \Omega}  
  \newcommand{\LichtPol} {\hat{\mathcal{E}}}  
\newcommand{\V}{V}
\newcommand{\CrossSec}{\sigma}

\newcommand{ \dip}{ \vec{\mu}}           %

\newcommand{\Huang}{X}

\newcommand{ \VibFr} { \omega}                       %

\newcommand{ \vc} { \kappa}                       %

\newcommand{\OOp}{D}
\newcommand{\OOpBar}{\bar{\OOp}}

\newcommand{\Bargman}{z}
\newcommand{\BargmanC}{\Bargman^*}

\newcommand{\HamAggEl}{H_{\rm el}}
\newcommand{\HamAggGes}{H}
\newcommand{\HamAggBath}{H_{\rm vib}}
\newcommand{\HamAggInt}{H_{\rm int}}

\newcommand{\AnzMon}{N}

\newcommand{\aDestroy}{a}
\newcommand{\aPlus}{a^{\dagger}}

\newcommand{\BathCor}{\alpha}

\newcommand{\kB}{k_B}

\newcommand{\e}{\mbox{e}}

\excludeversion{weg}

\newlength{\mylenunit}
\begin{document}
\title{Influence of complex exciton-phonon coupling on optical absorption and energy transfer of quantum aggregates} %

\author{Jan Roden and Alexander Eisfeld}
\affiliation{%
Max-Planck-Institut f\"{u}r
Physik komplexer Systeme,
N\"othnitzer Str.\ 38,
D-01187 Dresden, Germany
}%
\author{Wolfgang Wolff}
\affiliation{%
Marie-Curie-Gymnasium,
Giersbergstr.\ 39,
D-79199 Kirchzarten, Germany
}%
\author{Walter T.\ Strunz}
\affiliation{Institut f\"{u}r Theoretische Physik,
Technische Universit\"at Dresden, D-01062 Dresden, Germany}


\begin{abstract}
We present a theory that efficiently describes the quantum 
dynamics of an electronic excitation that is coupled to a 
continuous, highly structured phonon environment. 
Based on a stochastic approach to non-Markovian
open quantum systems, we develop a dynamical framework that 
allows to handle realistic systems
where a fully quantum treatment is desired yet
usual approximation schemes fail.
The capability of the method is demonstrated by calculating
spectra and energy transfer dynamics of mesoscopic molecular aggregates,
elucidating the transition from
fully coherent to incoherent transfer.

\end{abstract}

\maketitle
One often encounters the situation where a quantum particle or an elementary 
excitation couples to a complex environment. Examples range from the 
classical polaron problem \cite{polaron} 
over electron-phonon interaction in superconductors \cite{GrZe99_1296_}, 
ultracold impurity atoms immersed in a Bose–Einstein condensate and trapped in a tight optical lattice \cite{BrKlCl08_033015_},
molecular aggregates and crystals \cite{dye_aggr,ZhSp05_114701_,molaggr}, 
atoms or molecules in photonic band gap materials \cite{LaNiNi00_455_} 
or light-harvesting units in photosynthesis \cite{AmVaGr00__}.  
We speak of a quantum aggregate (QA), 
if the excitation can reside
on $N$ different sites, if it
can be handed over from site $n$ to
site $m$ mediated by a matrix element $V_{nm}$, and,
importantly, if the excitation couples to a complex environment.
In the following we will use the language of
exciton-phonon coupling in molecular aggregates, although this 
scenario is obviously much more general.

For {\it very small} QAs ($N<10$) it may be possible
to extract the few most relevant phonon modes and treat 
them fully quantum 
mechanically \cite{ZhSp05_114701_,Ke03_3320_}.
Then, however, the overall irreversible 
nature of the dynamics caused by the existence of many more 
environmental modes is ignored.
Other approaches, like the coherent exciton scattering approximation \cite{CES}, 
are best suited for large aggregates ($N\rightarrow \infty$), yet fail for
a small number of monomers.
If the overall influence of the vibrational environment is small, 
a perturbative approach
(Redfield) may be appropriate \cite{openquantsys, MaKue00__}. 

All these established approaches fail for the QAs we are interested in:
a quantum excitation that may reside on a
{\it finite} number of sites, significantly coupled to a 
{\it complex environment} of phonons that 
consists of a few distinct vibrations embedded in an overall dissipative bath.
We tackle this intricate regime of complex dynamics fully quantum
mechanically using ideas from a stochastic description of non-Markovian
open quantum systems \cite{DiSt97_569_} (see also \cite{openquantsys}). 
We present a dynamical framework based on a Stochastic Schr\"{o}dinger Equation 
(SSE) that allows us to determine the quantum dynamics 
emerging from a significant exciton-phonon coupling in complex QAs.
We are able to treat QAs consisting of
a very small to a large number of monomers, bridging a gap in previous 
approaches.  Moreover, being non-perturbatively, 
in a single unified theory we may describe 
exciton-dynamics ranging from fully coherent, over 
weakly perturbed, to strongly affected by the vibrations.
To demonstrate the capability of the method, it is applied
to study optical absorption and the coherent-incoherent transition of
energy transfer in ring-shaped molecular aggregates \cite{molaggr},
as they appear e.g.\
in the Light Harvesting units of some bacteria. 

We consider QAs where the wave functions of different monomers do not overlap
(tight binding) and each monomer has 
two electronic states with a transition 
energy $\ElTransE_n$ for monomer $n$.
A state in which monomer $n$ is electronically excited and all other 
monomers are in their electronic ground state is denoted by $\ket{\pi_n}$. 
 The Holstein model \cite{Ho59_325_}
includes the crucial influence of (possibly damped) vibrations
on each monomer and is given by the Hamiltonian  
\begin{equation}\label{Htotal}
\HamAggGes=\HamAggEl+\HamAggInt+\HamAggBath.
\end{equation} 
Here 
\begin{equation}
\HamAggEl=\sum_{n,m=1}^N\Big(\ElTransE_n\delta_{nm}+ 
\V_{nm}\Big)\ket{\pi_n}\bra{\pi_m}
\end{equation}
is the purely electronic part of the Hamiltonian and
\begin{equation}
\HamAggBath=\sum_{n=1}^{\AnzMon}\sum_j \hbar \VibFr_{nj} \aPlus_{nj}  
\aDestroy_{nj}
\end{equation}
describes the collection of phonon modes.
Here $\aDestroy_{nj}$ denotes the annihilation operator of 
mode $j$ of monomer $n$ with frequency $\VibFr_{nj}$.
For each monomer $n$, the sum over $j$ takes into account internal vibrations and their coupling to modes of the local environment.
The coupling of electronic excitation to these vibrations is contained in 
\begin{equation}
\HamAggInt=-\sum_{n=1}^{\AnzMon} \sum_j \vc_{nj} (\aPlus_{nj}  
+\aDestroy_{nj})\ket{\pi_n}\bra{\pi_n}
\end{equation}
where the coupling constants $\vc_{nj}$ are related to the dimensionless Huang-Rhys factor
$\Huang_{nj}$ through $\vc_{nj}=\hbar \VibFr_{nj} \sqrt{\Huang_{nj}}$
\cite{MeOs95__}.
An energy shift $\sum_j \hbar \VibFr_{nj} \Huang_{nj}$ is
incorporated into the transition energy $\ElTransE_n$. 

The complex structure of the phonon ``bath'' of monomer $n$ is encoded 
in the bath correlation function at temperature $T$
\cite{MaKue00__} 
\begin{equation}
\BathCor_n(\tau)= \int d\omega J_n(\omega)\Big( \cos(\omega \tau)
\coth\frac{\hbar \omega}{2 \kB T} - i \sin(\omega \tau) \Big) 
\end{equation}
with the spectral density
$J_n(\omega)=\sum_j |\vc_{nj}|^2 \delta (\omega-\omega_{nj})$
of monomer $n$
which is usually replaced 
by a smooth function to guarantee genuine irreversibility.
For simplicity we will restrict ourselves in the following to the zero temperature limit.

We use recently developed 
ideas from a SSE approach to open 
quantum system dynamics \cite{DiSt97_569_,DiGiSt98_1699_} 
to treat the model with a complicated continuous and structured phonon
distribution fully quantum mechanically.
In brevity, the SSE approach amounts to a
solution of the full Schr\"odinger equation for a total 
Hamiltonian of the type of Eq.~(\ref{Htotal}). It
may be derived using
a (Bargmann) coherent state basis~\cite{Ba61_187_} with $|z_{nj}\rangle = \exp(z_{nj}\aPlus_{nj})|0\rangle$ for each environmental (vibrational) degree of freedom.
Here $|0\rangle$ is the state where no vibrations are excited and $z_{nj}$ is a complex number. 
Thus (here for a zero
temperature environment) the full
state of system and environment at all times is written in the form \cite{Ba61_187_}
\begin{equation}\label{totalstate}
|\Psi(t)\rangle = \int\frac{d^2{\bf z}}{\pi}\ \e^{-|{\bf z}|^2}
|\psi(t,{\bf z}^*)\rangle |{\bf z}\rangle
\end{equation} 
with ${\bf z}$ representing the collection of coherent state labels 
$z_{nj}$. 
Remarkably, in an approximation to be discussed below
(see also section III.B of \cite{YuDiGi99_91_}),
the dynamics of the Holstein model (\ref{Htotal})
is now captured in the Schr\"odinger equation
\begin{eqnarray}
\label{Schr_final}
\partial_t |\psi(t,{\bf z}^*)\rangle
  & = &
-\frac{i}{\hbar} \HamAggEl  |\psi(t,{\bf z}^*)\rangle \\ \nonumber
& & -\sum_{m} |\pi_m\rangle\langle\pi_m|\big(\Bargman^{*}_{m}(t) - 
\OOpBar^{(m)}(t)\big)|\psi(t,{\bf z}^*)\rangle
\end{eqnarray}
in the small Hilbert space of the electronic degrees of freedom 
alone -- a huge reduction in complexity. In Eq.~(\ref{Schr_final}),
we use the abbreviations
$\Bargman^{*}_{m}(t)= -\frac{i}{\hbar}
\sum_j\kappa_{mj}z_{mj}^{*}\e^{i\omega_{mj}t}$
and
\begin{equation}
\label{def:OOpBar0}
\OOpBar^{(m)}(t)=\int_0^t\! d{s}\, \BathCor_{m}(t-s) \OOp^{(m)}(t,s).
\end{equation}
Here, $\OOp^{(m)}(t,s)$ represents a ${\bf z}^*$-independent operator 
in the electronic
Hilbert space, introduced to approximate a functional derivative
$\frac{\delta}{\delta  \BargmanC_{m}(s)} |\psi(t,{\bf z}^*)\rangle
\approx \OOp^{(m)}(t,s) |\psi(t,{\bf z}^*)\rangle$ 
that appears in the 
exact equation
\footnote{note that in Ref.\cite{DiSt97_569_,YuDiGi99_91_} this operator was named $O^{(m)}$.}.
The reasoning underlying this approximation is elaborated upon
further at the end of this Letter.
The operator $\OOp^{(m)}(t,s)$ is obtained by solving
\begin{equation}
\label{oop}
\begin{split}
\partial_t\OOp^{(m)}(t,s)
=&\Big[-\frac{i}{\hbar}\HamAggEl, \OOp^{(m)}(t,s)  \Big]\\
&+\sum_l\Big[ \ket{\pi_{l}}\bra{\pi_{l}}
  \OOpBar^{(l)}(t) ,  \OOp^{(m)}(t,s)\Big],
\end{split}
\end{equation}
with initial condition 
$\OOp^{(m)}(t=s,s)=-|\pi_m\rangle\langle\pi_m|$ \cite{YuDiGi99_91_}.

Equation (\ref{Schr_final}) (for the electronic state), together with 
Eqs.~(\ref{def:OOpBar0}) and (\ref{oop}) (for operators in the
electronic Hilbert space) is the new dynamical framework 
which will be used to determine all properties of interest of the QA
\footnote{For the numerical calculations of transfer dynamics we use the 
non-linear version of Eq.~(\ref{Schr_final})~\cite{YuDiGi99_91_}.}.
Together with expression (\ref{totalstate}),
$|\psi(t,{\bf z}^*)\rangle$ constitutes the full state and thus
all information about electronic and
vibronic degrees of freedom is available. 

The cross-section for absorption of light with 
frequency $\LichtFr$ in dipole-approximation at zero-temperature turns 
out to be connected to a simple autocorrelation function:
\begin{equation}
\label{app:CrossSec:c_nm}
\CrossSec(\LichtFr)=\frac{4 \pi}{\hbar c} \LichtFr\ 
\Real \int_{0}^{\infty}\!dt\ \e^{i  \LichtFr t}
\langle\psi(0,{\bf z=0})|\psi(t,{\bf z=0})\rangle.
\end{equation}
The state $|\psi(t,{\bf z=0})\rangle$ is obtained from 
Eq.~(\ref{Schr_final}) with initial condition
$
|\psi(0,{\bf z=0})\rangle=\sum_{n=1}^N (\LichtPol\cdot \dip_n)\ket{\pi_n}
$
where the geometry of the aggregate enters explicitly via the transition 
dipoles $\dip_n$ and the polarization of the light $\LichtPol$.
Note that only  the projection of the total state
onto the vibronic ground state $|0\rangle$ is needed, which is
the single solution of Eq.~(\ref{Schr_final}) with the choice 
$\Bargman^{*}_{m}(t)=0$.

More involved is the determination of
transport properties which requires the reduced density operator
$\rho(t) = $Tr$_{vib}|\Psi(t)\rangle\langle\Psi(t)|$. It is 
found by considering the $\Bargman^{*}_{m}(t)$ in
Eq.~(\ref{Schr_final}) to be independent 
colored stochastic processes with correlations
$\langle\!\langle z_m(t)z_n(s) \rangle\!\rangle=0$ and
$\langle\!\langle z_m(t) \rangle\!\rangle=0$.
The covariance of these processes is connected to the bath correlation 
function via
$\langle\!\langle z_m^*(t)z_n(s) \rangle\!\rangle
=\BathCor_{m}(t-s)\delta_{mn}$. 
It follows that
the reduced density operator of the electronic 
part can be obtained as an ensemble mean
$
\rho_t= \langle\!\langle 
|\psi(t,{\bf z}^*)\rangle\langle\psi(t,{\bf z}^*)|
\rangle\!\rangle
$
over the noises $z_1(t),\dots,z_N(t)$.

\begin{figure}
\includegraphics[width=0.6\mylenunit]{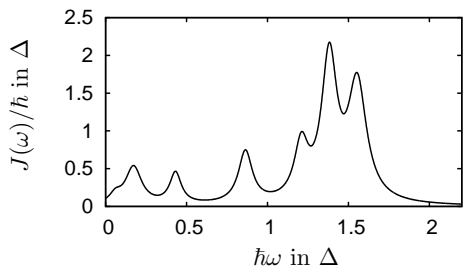}
\caption{\label{fig:spectrdensity}The spectral density used for
the calculation of spectra in Fig.~\ref{fig_monomer_and_j_band} and 
energy transfer in Fig.~\ref{fig:transfer}. The unit of energy is the width $\Delta$ of the resulting  monomer absorption spectrum (Fig.~\ref{fig_monomer_and_j_band}a).}
\end{figure}

\begin{figure}
\includegraphics[width=0.7\mylenunit]{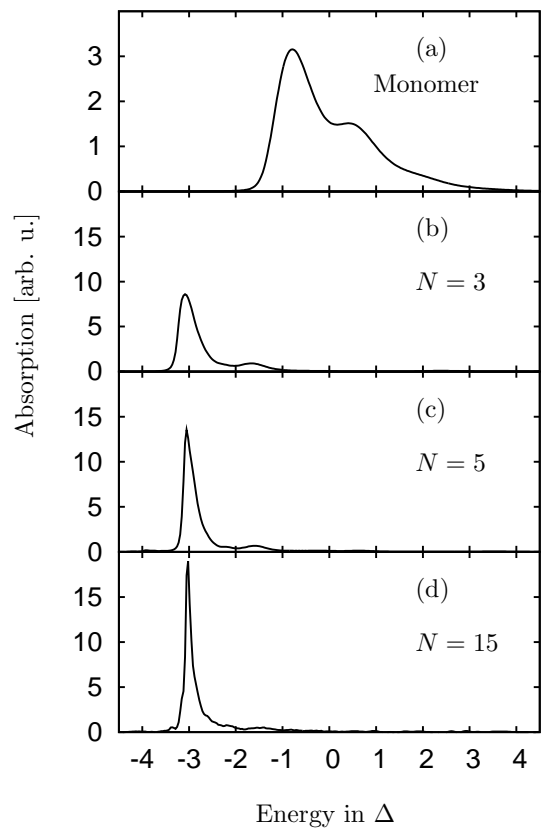}
\caption{\label{fig_monomer_and_j_band} (a) Absorption spectrum of the monomer (its width $\Delta$ (standard deviation) is used as the unit of energy). (b)-(d) J-band spectra of ring-shaped aggregates with $C=-2.6$~$\Delta$. The values of $N$ are indicated in the figures.}
\end{figure}

Our novel approach is now applied to
study optical absorption and transfer properties of molecular aggregates~\cite{molaggr}.
The spectral density of the monomers $J(\omega)$ is  taken to be a sum of
Lorentzians, see Fig.~\ref{fig:spectrdensity}, resulting in the monomer absorption spectrum shown in Fig.~\ref{fig_monomer_and_j_band}a, which nicely resembles that of a typical organic dye~\cite{ReWi97_7977_}.
In the following we will take the width $\Delta$ (standard deviation) of this monomer spectrum as the unit of energy (for organic dyes $\Delta$ is in the order of 0.1~eV).
The vibrational progression due to the high-energy modes with energies around 1.5~$\Delta$ in the spectral density (see Fig.~\ref{fig:spectrdensity}) is clearly visible. 
The considerable broadening of this progression mainly stems from the low-energy vibrations below 0.5~$\Delta$ in the spectral density.

In the following we focus on aggregates for which the absorption exhibits a narrow band, red-shifted w.r.t.\ the monomer absorption, the so-called J-band~\cite{molaggr,WaEiBr08_044505_}.
We consider an aggregate of $N$ identical monomers arranged equidistantly along a ring with transition dipoles lying in the plane of the ring, such that the angle between the transition dipoles of neighboring monomers is identical for all monomers.
In the calculations we have taken into account the interaction $V\equiv V_{n,n+1}$ between neighboring monomers only.
For the chosen geometrical arrangement, without coupling to vibrations, the aggregate absorption would be a single line, shifted by an energy $C\equiv 2V\cos(2\pi/N)$ w.r.t.\ the electronic monomer absorption line.

Fig.~\ref{fig_monomer_and_j_band}b-d shows aggregate absorption spectra for $C\! =\!-2.6$~$\Delta$ for different $N$. 
We find that the mean of the aggregate spectrum is shifted by the energy 
$C$ w.r.t.\ to the mean of the monomer spectrum, in accordance with sum 
rules \cite{BrHe70_1663_}. 
Furthermore, with increasing $N$ the 
vibrational structure vanishes and the lowest peak (around $-\!2.6$~$\Delta$) becomes narrower by roughly a factor $1/\sqrt{N}$.
This is the well-known effect of motional narrowing which leads to the 
narrow shape of the J-band of molecular aggregates
\cite{WaEiBr08_044505_,motion_narrow}, obtained here from a fully
dynamical calculation.
Upon increasing $|C|$ further the shape of the aggregate spectrum (especially the width) undergoes only very small changes, hardly noticeable even for $|C|\rightarrow \infty$.
Therefore one might assume that also other properties of the QA
will only slightly change when increasing $|C|$.  

\begin{figure}
\includegraphics[width=0.8\mylenunit]{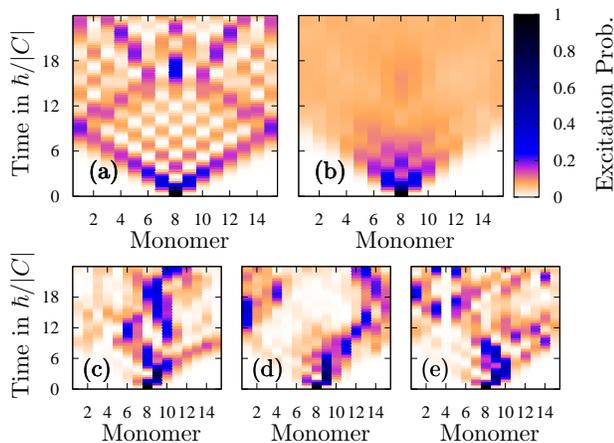}
\caption{\label{fig:transfer}Transfer of the electronic excitation energy on a ring-shaped 15-mer for $C=-2.6$~$\Delta$. 
Initially only monomer 8 is excited.
(a) Without coupling to a phonon bath.
(b) With coupling to a phonon bath with spectral density of Fig.~\ref{fig:spectrdensity}.
(c)-(e) Three of the 1000 single realisations over which the transfer in (b) is averaged.}
\end{figure}

However this is not the case as we will now show considering energy transfer
for the same situation as in Fig.~\ref{fig_monomer_and_j_band}d, (i.e.\
$N\!=\!15$ and $C\!=\!-2.6$~$\Delta$). 
Initially, the electronic excitation is chosen to be localised on monomer number eight. 
As unit of time we take the typical time $\hbar/|C|$ of intermonomer electronic excitation transfer~\cite{RoScEiBr_accepted}.
In Fig.~\ref{fig:transfer}b we show the time dependent probability to 
be electronically excited as a function of site number and time 
(note that the aggregate is ring-shaped).
For reference, in Fig.~\ref{fig:transfer}a the case where the electronic transfer does not couple to any vibrational modes is shown \cite{Me58_647_}. 
While the transfer in Fig.~\ref{fig:transfer}a, obtained from a purely 
electronic theory, exhibits clear excitation maxima over a long period of time, 
the excitation in Fig.~\ref{fig:transfer}b is distributed quickly over all 
monomers due to the coupling to the vibrational continuum 
(the transfer shown is averaged over 1000 realisations of the stochastic 
noise ${\bf z}^*$, but was well converged after only 600 realisations).
We have found that upon increasing $|C|$ the fast smearing of the excitation in Fig.~\ref{fig:transfer}b is suppressed and at about $C=-13$~$\Delta$ the purely electronic situation of Fig.~\ref{fig:transfer}a is reached.
This is quite remarkable, showing, that from the {\it width} of the J-band alone it is not easily possible to infer the influence of the phonon bath
on transfer properties.  
To gain deeper insight into the nature of the transfer, in Fig.~\ref{fig:transfer}c-e three of the 1000 single realisations over which the transfer 
in Fig.~\ref{fig:transfer}b is averaged are shown.
In these single realisations the excitation stays localised in a small region (about 3 monomers) and performs a random-walk-like motion. 

These considerations show that our dynamical framework
based on a SSE in the Hilbert space of electronic excitation allows an
efficient and
detailed description of properties of QAs, including complex vibrational
couplings.
Therefore an examination of the approximation underlying
Eq.\ (\ref{Schr_final}) is in order.
It is based on a functional expansion of $\frac{\delta}{\delta  \BargmanC_{m}(s)} |\psi(t,{\bf z}^*)\rangle$ w.r.t. the noise ${\bf z}^*$ \cite{YuDiGi99_91_}, taking only the lowest order term into account.
This approximation  has been confirmed to be true in many cases of interest:
it is true near the Markov limit (Lindblad), and contains the weak
coupling (Redfield) limit~\cite{VeAlGaSt05_124106_}.
Moreover, it holds true for many soluble
cases, including the case of independent monomers ($V_{nm}=0$) of this Holstein
model.
To check the quality of the approximation beyond the usual limits (Markov, Redfield), we investigated the case of a spectral density consisting of a single Lorentzian in more detail.
For the dimer ($N=2$) we were able to compare with spectra obtained from full quantum calculations and found overall good agreement.
Last but not least, we have confirmed that the sum 
rules \cite{BrHe70_1663_,He77_1795_} for the first five moments of the
absorption spectrum are satisfied.

To conclude, we have developed a new dynamical framework for the determination
of optical and transport
properties of QA. This method allows a fully quantum treatment
with realistic complex vibrational environments.
The usefulness has been shown by considering the emergence of the J-band 
as $N$ grows and
by capturing the transition from coherent to incoherent energy transfer.
The next step is to investigate the much more complicated case of the H-band
\cite{EiBr06_376_} where the details of the vibrational structure play a more
pronounced role.
Since the method is based on the time-propagation of a SSE, it is also ideally
suited to include external time-dependent fields.
This should enable us to efficiently study coherent control schemes or 
describe multidimensional spectroscopy using realistic spectral densities.
Clearly, as indicated in the introduction, the model we solve here and
variants thereof appear in many applications well beyond molecular aggregates 
which will be a subject of future research.

\begin{acknowledgments}
We thank John S. Briggs for many fruitful discussions and
for initiating this collaboration.
\end{acknowledgments}

\end{document}